# Installation, commissioning, and testing of the HB650 CM at PIP2IT

M White[1], J Makara[1], S Ranpariya[1], L Pei[1], M Barba[1], J Subedi[1], J Dong[1], B Hansen[1], A E T Akintola[1,2], J Holzbauer[1], J Ozelis[1], S Chandrasekaran[1], V Roger[1]

[1]Fermi National Accelerator Laboratory, PO Box 500, Batavia, IL 60510
[2]Accelerator Science and Technology Centre, STFC Daresbury Laboratory, Keckwick Lane, Warrington, WA4 4AD, UK

Email: mjwhite@fnal.gov

**Abstract**. The Proton Improvement Plan-II (PIP-II) is a major upgrade to the Fermilab accelerator complex, featuring a new 800-MeV Superconducting Radio-Frequency (SRF) linear accelerator (LINAC) powering the accelerator complex to provide the world's most intense high-energy neutrino beam. This paper describes the conversion of the PIP-II Injector Test Facility (PIP2IT) cryogenic system into a test stand for PIP-II High-Beta 650 MHz (HB650) cryomodules at Fermilab's Cryomodule Test Facility (CMTF). A description of the associated mechanical, electrical, and controls modifications necessary for testing HB650 cryomodules are provided. The cooldown and warmup requirements, procedures and associated controls logic is described.

## 1. Introduction and System Overview

The Cryomodule Test Facility (CMTF) at Fermilab has two caves [1] for testing and commissioning Superconducting Radio-Frequency (SRF) cryomodules for the Linac Coherent Light Source II High Energy (LCLSII HE) upgrade and Proton Improvement Plan-II (PIP-II) projects. This paper specifically describes the conversion of the PIP-II Injector Test Facility (PIP2IT) cryogenic system into a test stand for PIP-II High-Beta 650 MHz (HB650) cryomodules at CMTF.

There are 5 Mycom 2016C screw compressors available at CMTF, with four compressors available for running the superfluid cryogenic plant (SCP) and one compressor used for purifying all subatmospheric return flow through a liquid nitrogen cooled charcoal adsorber. Each compressor has a 300 kW (400 hp) motor and delivers up to 60 g/s with a suction pressure of approximately 1 bar and a discharge pressure of up to 20 bar [2].

The measured capacity of the SCP in liquefier mode is 720 W on the high temperature thermal shield (HTTS) circuit 118 W on the low temperature thermal intercept (LTTI) circuit while providing 25 g/s of liquid helium [3]. The SCP cold compressors have not been reliable; therefore, maintaining the 30 mbar (23 torr) bath pressure for the superconducting radio-frequency (SRF) cavities to operate at 2.0 K has been accomplished solely through warm vacuum pumping. Without the cold compressors the SCP still has enough capacity that both test stands can operate in any mode independent of the other test stand. The only exception is that both test stands can not perform a full unit test with all SRF cavities at maximum RF power simultaneously. The repair of the nitrogen precooling heat exchanger is described in a companion paper at this conference [4].

There are three warm vacuum pump skids. The first skid is comprised of a Kinney KMBD 10000 roots blower and a Kinney KLRC 2100 liquid ring pump [2] and is typically used for LCLSII HE cryomodule testing, but there is a bypass that allows the first skid to be used for PIP2IT testing if needed. The second and third warm vacuum pumping skids use Kinney KMBD 3200 roots blower and Kinney KLRC 950 liquid ring pumps [5].

The PIP-II project includes five different types of SRF cryomodules [6]. The first type of SRF cryomodule in the PIP-II linac in the beam-direction is the Half-Wave Resonator (HWR) cryomodule, which includes 8 SRF cavities operating at 162.5 MHz and 8 superconducting cavities. The second type of SRF cryomodule in the beam-direction is a Single Spoke Resonator (SSR1) cryomodule, which includes 8 SRF spoke cavities operating at 325 MHz and 8 superconducting magnets. The first use of the PIP-II Injector Test Facility (PIP2IT) cave was to commission the front end of the PIP-II linac, which included running beam through both the HWR and prototype SSR1 cryomodule [7]. Moving forward the PIP2IT cave will no longer be used to accelerate beam, but will instead be used as a cryomodule test stand to test and commission each of the PIP-II cryomodules. Results from testing the prototype SSR1 cryomodule have already been reported [8-9]. Most recently, the PIP2IT cave was used for testing the prototype High-Beta 650 MHz (HB650) cryomodule [10-12], which houses 6 elliptical-shaped SRF cavities operating at 650 MHz

## 2. Installation

### 2.1. Mechanical

The placement order of the cryogenic bayonets on PIP-II cryomodules was standardized after the HWR cryomodule was already in production. Therefore, HWR cryomodule has a unique arrangement of bayonets and U-tube connections relative to all other PIP-II cryomodules. The original transfer line for the PIP2IT accelerator [13] had bayonets arranged specifically for the HWR cryomodule on the bayonet can. The largest mechanical modification to the PIP2IT cryogenic system as part of the conversion from an accelerator to a cryomodule test stand was to build the adapter transfer line shown in figures 1 to put the bayonets in the right order to avoid u-tubes crossing over each other. Figure 2 shows the adapter transfer line installed under the platform used to access U-tubes. All of the HWR U-tubes had to be modified to the new vertical and horizontal dimensions necessary to connect to the HB650 cryomodule.

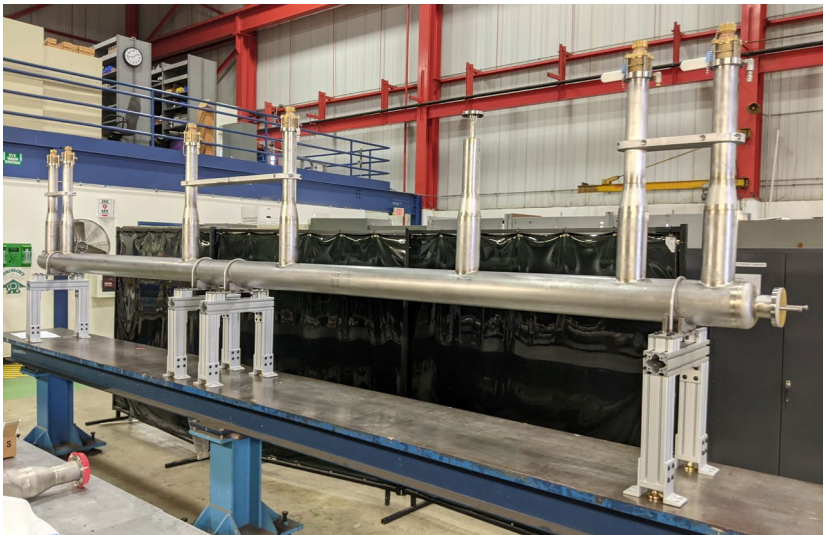

**Figure 1.** Adapter transfer line used to accommodate HB650 cryomodule U-tubes prior to installation in the PIP2IT cave

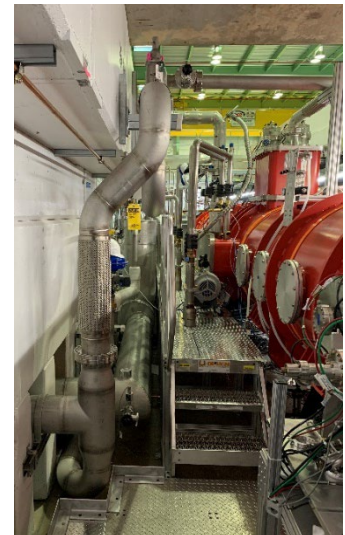

**Figure 2**. Adapter transfer line as-installed

The other major mechanical change to the cryogenic system was to the pressure relief system. The SRF cavity circuit for the HWR cryomodule was designed for 2 bar absolute, which resulted in a 1 bar gauge rupture disk that vented into the PIP2IT. The penetrations in the PIP2IT cave are limited in size and involve multiple bends to prevent radiation hazards outside the cave, so for the HWR cryomodule the pressure drop was too restrictive to vent the rupture disk outside the cave. The SRF cavities for the SSR1 and HB650 cryomodules are designed for 2.05 bar absolute above 80 K and 4.10 bar absolute below 80 K. The rupture disk for HWR was replaced with a new rupture disk set to burst at just under 3 bar.g. The higher burst pressure corresponding larger fluid density made it practical to connect the rupture disk to a vent line that discharges outside the PIP2IT cave, which is desirable to lower the oxygen deficiency hazard in the PIP2IT cave.

*2.2. Process control valves*
The PIP2IT transfer line control valves regulating processes for the HB650 cryomodule are shown in Figure 3. The HTTS supply is nominally at 40 K and the return is nominally at 80 K. The PIP2IT bayonet can includes a warm gas mixing control valve utilizing compressor discharge that allows the supply temperature to be controlled anywhere between 290 K and 40 K. The warm gas control valve regulates mixing pressure and the 40K supply valve regulates mixing temperature during cooldown and warmup. The 80K return valve in the PIP2IT transfer line is used to regulate the HTTS return temperature during normal operation. There is also a vent valve to compressor suction in the PIP2IT transfer line, with the vent valve controlled on a measured flow rate to avoid excessive ice build-up on the compressor suction piping.

The transfer line has supercritical helium supply nominally at 5 K that is in the range of 3.0 to 3.5 bara. The PIP2IT bayonet can similarly has a warm gas mixing control valve that allows the supply temperature to the SRF cavities to be controlled anywhere between 290 K and 5 K during a warmup or cooldown. After passing through the U-tube, the 5 K supply flow then branches off in two directions within the HB650 cryomodule. The first branch is the LTTI, which is used to intercept heat conducted towards the 2.0 K SRF cavity bath. The second branch of the 5 K supply is used for cooling the SRF cavities through the PVCD control valve and also maintaining liquid level in the 2.0 K SRF cavity bath using the PVJT control valve. The HB650 cryomodule PVJT valve is controlled following the model used for LCLSII cryomodule testing [14]. During normal operation the 8K return valve in the PIP2IT can regulates the LTTI return temperature and the 2K return valve in the PIP2IT can regulates the SRF cavity bath pressure.

The LTTI return is sent to compressor suction during a cooldown or warmup. The SRF cavity bath return also must be sent to compressor suction during cooldown, since there is concentric piping with the 5K supply in the PIP2IT transfer line. Avoiding heat exchange with cooldown return flow in the concentric piping is important to ensuring a smooth and stable cooldown supply conditions. During normal operation at 2.0 K the SRF cavity bath return flow is routed through the concentric transfer line, which could recapture refrigeration in addition to the refrigeration recovered within the HB650 cryomodule Joule-Thomson (JT) heat exchanger. After the section of concentric transfer line, the SRF cavity bath return flow breaks out of the insulating vacuum jacket, the flow is warmed by an immersion heater, and then either flows through the warm vacuum pump(s) or bypasses the warm vacuum pumps into compressor suction.

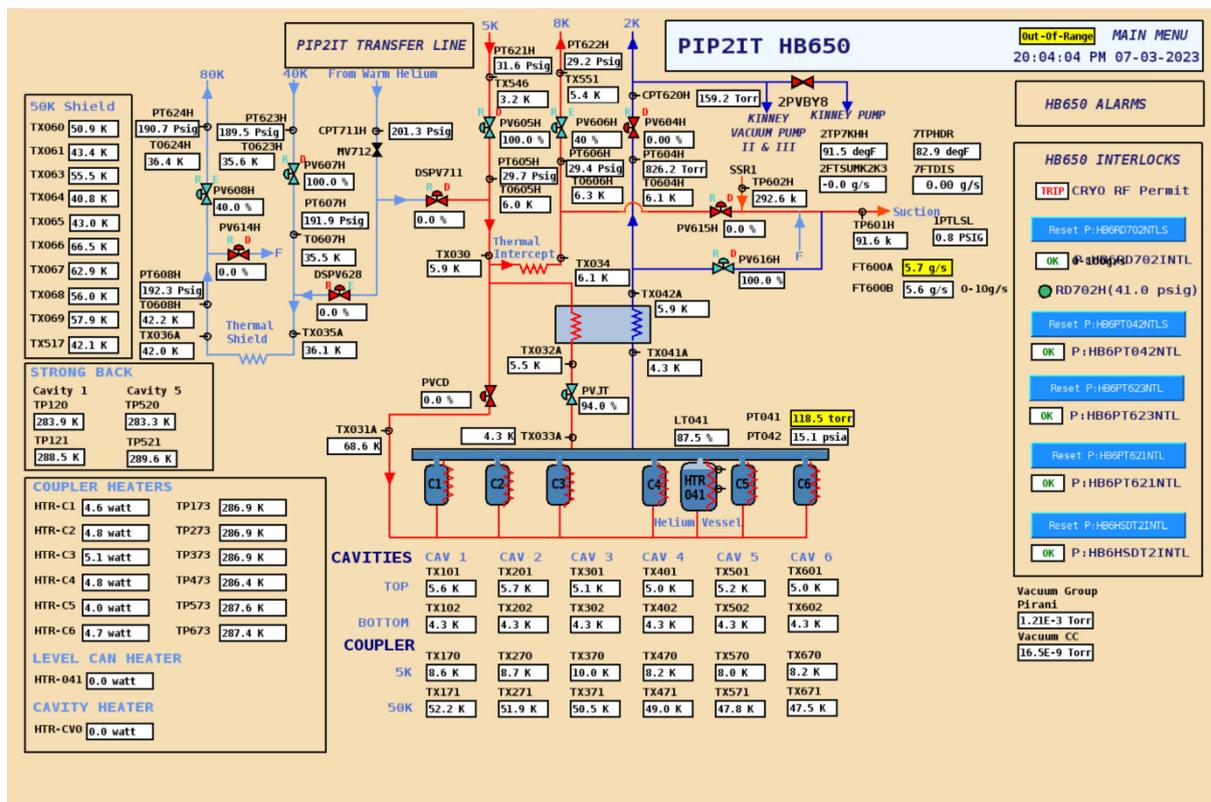

**Figure 3.** The Synoptic HMI for the PIP2IT HB650 Cryomodule showing the arrangement of control valves and instrumentation as the cryomodule was initially filled with liquid helium. The SRF cavity bath is being vented to compressor suction. The interlocks and associated reset buttons are shown on the right.

### 2.3. Instrumentation

The multi-level distribution control system uses Siemens S7-400 PLC as its central control. The S7-400 PLC measures and controls two pressure transducers, two flow meters, ten platinum RTD temperature sensors, 31 heater readbacks, and one level sensor that were added for HB650 cryomodule testing. Further details on the control system are described elsewhere [15-16].

Four of these Platinum RDS sensors are on the strongback and the other six Platinum RTD sensors are located on the RF coupler ceramic window flanges. Four 10W heaters are connected in parallel on each coupler outer conductor. Each coupler heater regulates on the coupler window platinum RTD and keeps it above the dew point in the event that the coupler air supply is lost.

There are 81 Lakeshore Cernox RTD temperature sensors on the HB650 cryomodule to monitor the temperature of the cavity. Sensors are mounted on the process piping adjacent to the bayonets to monitor the inlet and outlet temperatures of each process flow. There are also temperature sensors on each port on the JT heat exchanger, downstream of the PVJT and PVCD valves, and on the SRF cavity relief piping thermal intercepts. Each cavity has temperature sensor mounted on the top and bottom as well as at the 5K and 50K thermal intercept locations on RF couplers. Two of the bottom G10 supports have sensors at the 5K and 50K thermal intercept locations. The remaining sensors are installed on the HTTS and supports internal to the top hat of the HB650 cryomodule.

The cooldown valve (PVCD) and the Joule-Thomson (PVJT) valve are controlled through the Siemens SIPART PS2 positioner which communicates to the PLC over Profibus communication and controlled by the operator through the EPICS HMI. All other PIP2IT cryogenic distribution system control valves are controlled by the operator through ACNET and Synoptic displays.

There are a total 31 heaters on cryomodule including the coupler heaters previously mentioned. There is a 100W heater HTR-041 located on helium container mounted in the SRF cavity bath level can. This heater has a four-wire connection to the electrical feedthrough on the vacuum vessel, so the heater power supply controls power level based on heat dissipated in the element. Six heaters are connected in parallel with a total capacity of 300W and are mounted on the cavity helium vessel. The cavity heaters have a two-wire connection to the electrical feedthroughs on the vacuum vessel, so the heater power supply controls the power level based on the heat dissipated in the heater element and the wiring internal to the vacuum vessel.

**3. Commissioning**

*3.1. Controlled Cooldown*
The HTTS supply has a warm gas mixing control valve that allows the supply temperature to be controlled anywhere between 290 K and 40 K. There are no direct cooldown or warmup rate constraints on thermal shield. The HTTS thermal shield is designed for nominal cooldown or warmup rate of 20 K/hr. However, there is maximum allowable temperature difference of 100K between any two of the following four sensors: TX-035 (40K supply), TX-036 (80K return), TX-060 (near base of top hat shield) and TX-063 (also near base of top hat shield)

Similarly, the supply to the LTTI and SRF cavity circuits also has a warm gas mixing control valve that allows the supply temperature to be controlled anywhere between 290K and 5K. There are no thermal constraints on SRF cavity circuit when cavity temperatures are greater than 175 K or less than 90 K. However, SRF cavities should be cooled at least 20K/hr between 175 K and 90 K to avoid Q-disease. There are no thermal constraints on LTTS cooldown or warmup.

One desired future improvement would be to install an immersion heater on the compressor suction return piping, so that flow rates greater than 20 g/s could be used during cooldown without excessive ice build-up. An additional desired future improvement would be to generate a Sequential Function Controller to perform the cooldown and warmup steps while maintaining temperature constraints.

*3.2. Interlocks & Alarms*
A total of five interlocks were used for the HB650 CM cryogenic operations at PIP2IT. The first two interlocks are intended to prevent overpressure conditions and to conserve helium in the event overpressure conditions do occur. To prevent overpressure, the 5K/2K circuit valves venting the cryomodule to compressor suction open and all other 5K/2K circuit cryogenic valves close if 2K bath pressure exceeds 1.7 bar (25 psia). To mitigate the loss of helium, all control valves on 5K/2K circuit close if the rupture disk bursts. There is approximately 530 L of liquid helium in the HB650 cryomodule during operation.

The rupture disk interlock takes priority over all other interlocks. There are no contradictions in the actions of the other four interlocks, so those four interlocks all have an equal lower priority level. All interlocks require the signal to be outside of the interlock threshold for two seconds before initiating the trip action to avoid nuisance trips due to signal noise.

The third and fourth interlocks are intended to prevent accidentally sending warm gas backwards into the cryogenic transfer line. If the HTTS warm mixing valve is open and the transfer line supply pressure is less than the mixing pressure, then the interlock closes all HTTS cryogenic valves and open the control valve to compressor suction. If the SRF cavity warm mixing valve is open and the transfer line pressure is less than the mixing pressure, then the interlock closes all SRF cavity circuit cryogenic valves and open the valve to compressor suction. Using this logic warm gas can flow backwards into the transfer line for no more than 2 seconds.

The fifth interlock is intended to prevent allowable thermal stresses in the HTTS shield from being exceeded. If the HTTS 100 K temperature constraint is exceeded, then close all HTTS cryogenic valves and open the valve to compressor suction. With no flow through the thermal shield the temperature distribution should equalize over time. Once the four HTTS temperatures are back within the 100K temperature difference constraint the operator can restart the cooldown or warmup.

*3.3. Fast Cooldown*

SRF cavities soak to 50K for at least 2 hours prior to start of fast cooldown. During fast cooldowns the SCP supply capacity is supplemented by rapidly boiling off liquid in the 3,000 L LHe dewar using the maximum allowable power on the heater. Based on LCLSII cryomodule testing, which uses a cryogenic temperature Coriolis flowmeter in the 5K supply line, it is known the SCP can supply in excess of 99 g/s for short durations of less than 10 minutes. PIP-II cryomodules have a similar requirement for a fast cooldown of 20 K/min through the niobium superconducting transition temperature of 9.2 K. Based on tests of individual HB650 SRF cavities it was expected that at least 80 g/s would be needed to ensure the cooldown requirements were satisfied.

Nitrogen doped niobium RF cavities can very efficiently store energy, producing large beam acceleration for low heat loads. This efficiency can be significantly degraded by magnetic fields trapped in the bulk superconductor. These trapped fields are caused by slow/quasi-isothermal cooling through superconducting transition in the presence of ambient magnetic fields. In addition to significant design effort spent to minimize ambient fields, large and directional thermal gradients across the superconductor during transition has the effect of sweeping the magnetic fields out of the bulk material, more closely approximating the true Meissner state. This thermal gradient is achieved by using very high cryogen flow rates directed, by design, at one side of the cavities, exhausting at the opposite end. Prototype tests of these designs will include temperature sensors in the helium space directly on the cavities in addition to sensors in the insulating vacuum on the outside of the cavity helium jackets. In this way temperature profiles at different flow rates on the helium vessels can be correlated with the direct cavity thermal profile [17].

The PIP2IT cryogenic distribution system does not include any cryogenic temperature flowmeters, so in real-time the supply flow rate to the SRF cavities is unknown during fast cooldown. However, knowing the inlet temperature, inlet pressure, valve position, and outlet pressure the flow rate through the PVCD control valve can be estimated using the valve sizing equations in IEC 60534. PVCD uses a $C_v$ = 3.2 equal percentage bullet with a 100:1 ratio. The estimated flow rate through PVCD based on archived data was calculated to be in excess of 150 g/s. The cooldown rate at the bottom of the SRF cavities greatly exceeded 20 K/min and the cooldown rate at the top of the cavities remained close to 20K/min approaching the niobium transition temperature of 9.2K. A plot of the fast cooldown is shown in figure 4. The cooldown rates at the top and bottom of each cavity were similar across the string of cavities.

## 4. Conclusion and Future Testing Plans

Currently the prototype HB650 cryomodule is warm. Several activities are taking place. First, the stainless-steel stems for control valve PVCD and PVJT have been replaced with G10 stems. Secondly, the cryomodule side port has been opened to inspect for damaged MLI, thermal shorts, or other anomalous source of heat load, but no anomalies were found. Lastly, the RF coupler ports are being disassembled to inspect MLI and as necessary install additional MLI to prevent radiation from passing through the gap between the RF coupler flange and the high temperature thermal shield, with radiation passing through this gap appearing to the most likely cause of higher-than-expected heat loads. The plan is to then cooldown the prototype HB650 again to measure improvements in heat load performance.

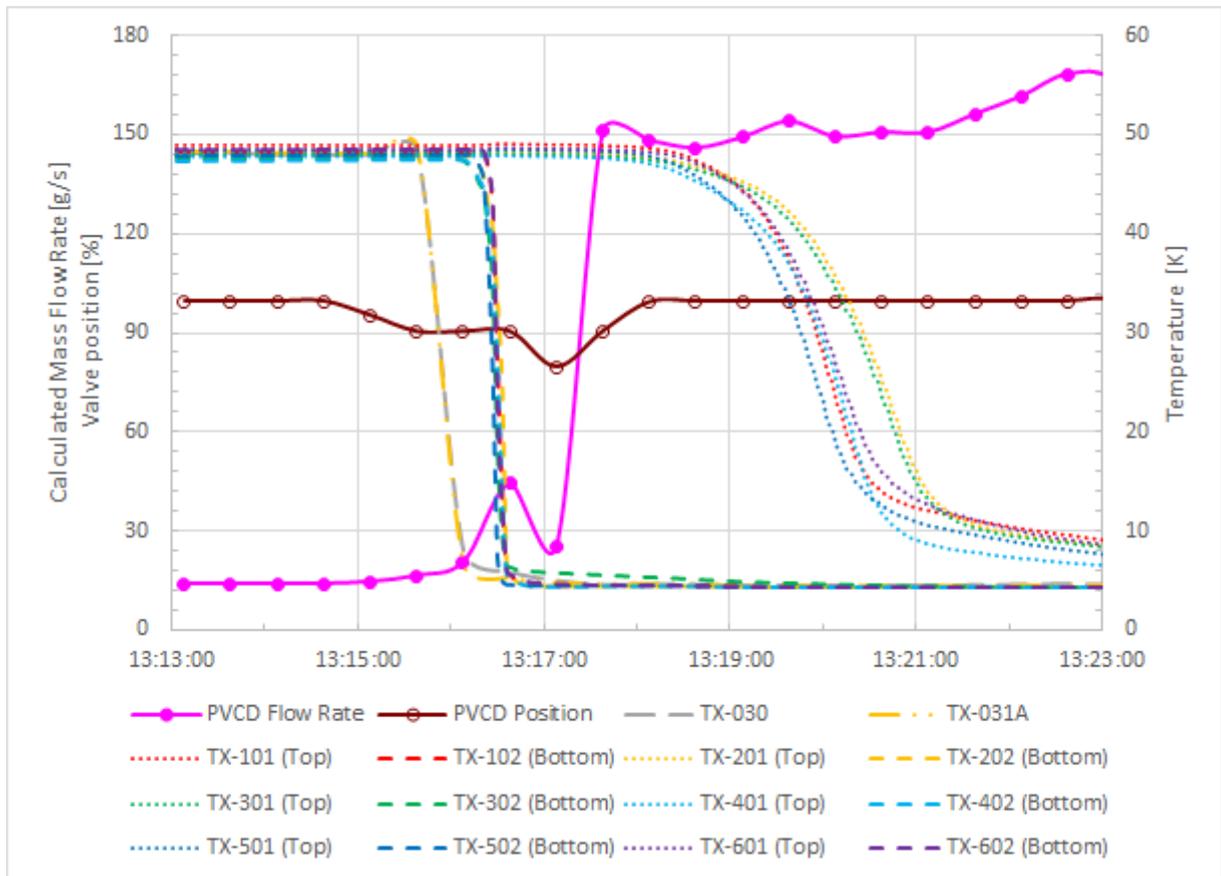

**Figure 4.** Plot showing the fast cooldown of the HB650 Cryomodule. TX-030 is upstream of PVCD and TX-031 is downstream of PVCD. The first digit in temperature sensor number corresponds to the cavity number in TX-101 through TX-602.

The next step is to ship the prototype HB650 CM to STFC and back to Fermilab [18-19]. This requires removing and re-installing the top hat. This provides another opportunity for inspection, adding additional instrumentation, and/or making modifications to internal piping. The plan is to then cooldown the prototype HB650 yet again to see if there are any changes in heat load performance or if there was any damage that occurred during transportation and handling.

Finally, there will be a retest the prototype SSR1 cryomodule now that a warm flow meter is available for measuring HTTS and LTTI flow. The HTTS and LTTI heat load measurements for the prototype SSR1 cryomodule had a high uncertainty since no flow meter was available. Additionally, the inlet and outlet temperatures will be controlled to more closely match the intended design temperatures for the PIP-II transfer line. The prototype SSR1 cryomodule can be cooled down independently of the HB650 cryomodule,

**Acknowledgments**

This manuscript has been authored by Fermi Research Alliance, LLC under Contract No. DE-AC02-07CH11359 with the U.S. Department of Energy, Office of Science, Office of High Energy Physics. The authors wish to recognize the dedication and skills of APS-TD/Cryogenics technical staff involved in the installation and commissioning of this system.